\documentclass[runningheads]{llncs}
\usepackage{hyperref}
\usepackage{subcaption}
\usepackage{tabularx}
\usepackage{listings}
\usepackage{verbatim}
\usepackage[table]{xcolor}
\usepackage{setspace}
\usepackage{enumitem}
\usepackage{float}
\usepackage[]{caption}
\usepackage{orcidlink}
\usepackage{pdfpages}
\usepackage{enumitem}
\usepackage{graphicx}
\usepackage{booktabs}
\usepackage{array}
\begin{document}
\title{Evaluating Large Language Models in Process Mining: Capabilities, Benchmarks, and Evaluation Strategies}
\titlerunning{Evaluating Large Language Models in Process Mining}

\author{
Alessandro Berti\inst{1,2}\orcidlink{0000-0002-3279-4795},
Humam Kourani\inst{1,2}\orcidlink{0000-0003-2375-2152},
Hannes Häfke\inst{1}\orcidlink{0000-0002-2845-3998},
Chiao-Yun Li\inst{1,2}\orcidlink{0009-0002-3767-7915},
Daniel Schuster\inst{1,2}\orcidlink{0000-0002-6512-9580}
}
\authorrunning{A. Berti et al.}
\institute{Fraunhofer FIT, Sankt Augustin, Germany  \and
Process and Data Science Chair, RWTH Aachen University, Aachen, Germany \\
\email{\{alessandro.berti,humam.kourani, hannes.haefke, chiao-yun.li, daniel.schuster\}@fit.fraunhofer.de}}

\maketitle

\begin{abstract}
Using Large Language Models (LLMs) for Process Mining (PM) tasks is becoming increasingly essential, and initial approaches yield promising results.
However, little attention has been given to developing strategies for evaluating and benchmarking the utility of incorporating LLMs into PM tasks.
This paper reviews the current implementations of LLMs in PM and reflects on three different questions.
1) What is the minimal set of capabilities required for PM on LLMs? 2) Which benchmark strategies help choose optimal LLMs for PM? 3) How do we evaluate the output of LLMs on specific PM tasks?
The answer to these questions is fundamental to the development of comprehensive process mining benchmarks on LLMs covering different tasks and implementation paradigms.
\end{abstract}

\renewcommand{\sectionautorefname}{Section}
\renewcommand{\subsectionautorefname}{Section}
\renewcommand{\subsubsectionautorefname}{Section}

\section{Introduction}
\label{sec:introduction}

\begin{figure*}[!t]
\centering
\includegraphics[width=0.8\textwidth]{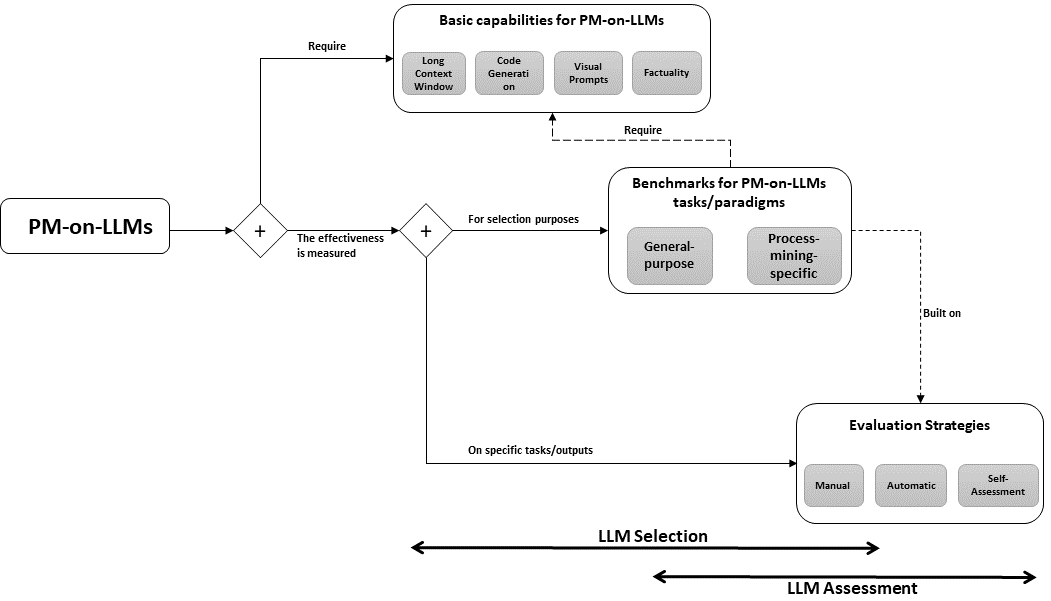}
\caption{Outline of the contributions of this paper.}
\label{fig:outline}
\vspace{-3mm}
\end{figure*}

Process mining (PM) is a data science field focusing on deriving insights about business process executions from event data recorded by information systems \cite{DBLP:books/sp/Aalst16}. Several types of PM exist, including
\emph{process discovery} (learning process models from event data), \emph{conformance checking} (comparing event data with process models), and \emph{process enhancement} (adding frequency/performance metrics to process models).
Although many automated methods exist for PM, human analysts usually handle process analysis due to the need for domain knowledge. Recently, LLMs have emerged as conversational interfaces trained on extensive data \cite{DBLP:journals/bise/TeubnerFWAH23}, achieving near-human performance in various general tasks \cite{DBLP:conf/iclr/ZhouMHPPCB23}. Their potential in PM lies in
\emph{embedded domain knowledge} useful for generating database queries and insights \cite{DBLP:conf/emnlp/PetroniRRLBWM19}, \emph{logical and temporal reasoning capabilities} \cite{DBLP:journals/corr/abs-2302-04023,DBLP:journals/corr/abs-2304-03439},
\emph{inference abilities over structured data} \cite{DBLP:conf/emnlp/JiangZDYZW23}.
Prior research has asserted the usage of LLMs for PM tasks \cite{DBLP:journals/corr/abs-2307-12701,DBLP:journals/corr/abs-2307-02194}. However, a comprehensive discussion on necessary capabilities for PM, LLMs' suitability evaluation for process analytics, and assessment of LLMs' outputs in the PM context is lacking.

The three main contributions of this paper are summarized in Fig. \ref{fig:outline}. First, building upon prior work \cite{DBLP:journals/corr/abs-2307-12701,DBLP:journals/corr/abs-2307-02194} proposing textual abstractions of process mining artifacts and an experimental evaluation of LLMs' responses, the essential capabilities that LLMs must have for PM tasks are derived in \autoref{sec:llmsCapabilitiesForPm}. The aforementioned capabilities allow us to narrow down the field of LLMs to those that meet these requirements. Next, evaluation benchmarks for selecting suitable LLMs are introduced in \autoref{sec:relevantLlmsBenchmarks}, incorporating both process-mining-specific and general criteria such as reasoning, visual understanding, factuality, and trustworthiness. Finally, we suggest automatic, human, and self-assessment methods for evaluating LLMs' outputs on specific tasks in \autoref{sec:llmsOutputsEvaluation}, aiming to establish a comprehensive PM benchmark and enhance confidence in LLMs' usage, addressing potential issues like hallucination.

This paper provides an orientation to process mining researchers investigating the usage of LLMs, i.e., this paper aims to facilitate PM-on-LLMs research.

\section{Background}
\label{sec:background}

LLMs enhance PM with superior capabilities, handling complex tasks through data understanding and natural language processing. This section covers PM tasks with LLMs (\autoref{sec:pmTasksLlms}) and the adopted implementation paradigms (\autoref{sec:implStrategiesPMonLLMs}) along with the provision of additional domain knowledge.

\vspace{-2mm}
\subsection{Process Mining Tasks for LLMs}
\label{sec:pmTasksLlms}

This subsection explores a range of PM tasks in which LLMs have already been adopted for process mining research. LLMs facilitate the automation of generating \emph{textual descriptions} from process data, handling inputs such as event logs or formal process models \cite{DBLP:journals/corr/abs-2307-02194}. They also generate \emph{process models} from textual descriptions, with studies showing LLMs creating BPMN models and declarative constraints from text \cite{10.1007/978-3-031-50974-2_34}. In the realm of anomaly detection, LLMs play a crucial role in identifying process data \emph{anomalies}, including unusual activities and performance bottlenecks, offering context-aware detection that adapts to new patterns through prompt engineering. This improves versatility over traditional methods \cite{DBLP:journals/corr/abs-2307-12701,DBLP:journals/corr/abs-2307-02194}. For \emph{root cause analysis}, LLMs analyze event logs to suggest causes of anomalies or inefficiencies, linking delays to specific conditions or events. This goes beyond predefined logic, employing language processing for context-aware analysis \cite{DBLP:journals/corr/abs-2307-12701,DBLP:journals/corr/abs-2307-02194} In ensuring \emph{fairness}, LLMs identify and mitigate bias in processes, suggesting adjustments. They analyze processes like recruitment to detect disparities in rejection rates or delays by gender or nationality, aiding in fair decision-making \cite{fairness,DBLP:journals/corr/abs-2307-02194}. LLMs can also interpret and explain \emph{visual data}, including complex visualizations, by describing event flows in dotted charts and identifying specific patterns, such as batch processing. For process improvement, after PM tasks identify and analyze problems, LLMs can suggest actions and propose new process constraints \cite{fairness,DBLP:journals/corr/abs-2307-02194}.

\vspace{-2mm}
\subsection{Implementation Paradigms of Process Mining on LLMs}
\label{sec:implStrategiesPMonLLMs}

To effectively employ LLMs for PM tasks, specific implementation paradigms are required \cite{DBLP:journals/corr/abs-2307-12701,DBLP:journals/corr/abs-2307-02194}. This section outlines key approaches for implementing LLMs in PM tasks. We distinguish three main strategies:
\begin{itemize}
\item \noindent \emph{Direct provision of insights}: A prompt is generated that merges data abstractions with a query about the task.  Also, interactive dialogue between the LLM and the user is possible for step-by-step analysis. The user starts with a query and refines or adjusts it based on the LLM's feedback, continuing until achieving the desired detail or accuracy, such as pinpointing process inefficiencies. For instance, to have LLMs identifying unusual behavior in an event log, we combine a textual abstraction of the log (such as the directly-follows graph or list of process variants) with a question like ``Can you analyze the log to detect any unusual behavior?''
\item \noindent \emph{Code generation}: LLMs can be used to create structured queries, like SQL, for advanced PM tasks \cite{DBLP:journals/corr/abs-2307-09909}. Rather than directly asking LLMs for answers, users command LLMs to craft database queries from natural language. These queries are then executed on the databases holding PM information. It is applicable to PM tasks that can be converted into database queries, such as filtering event logs or computing the average duration of process steps. Also, LLMs can be used to generate executable programs that use existing PM libraries to infer insights over the event data \cite{DBLP:conf/er/Harer23}.
\item \noindent \emph{Automated hypotheses generation}: Combining the previous strategies by using textual data abstraction to prompt LLMs for autonomous hypotheses generation \cite{DBLP:journals/corr/abs-2307-12701,DBLP:journals/corr/abs-2307-02194}. The hypotheses are accompanied by SQL queries for verification against event data. Results confirm or refute these hypotheses, with potential for LLM-suggested refinements of hypotheses.
\end{itemize}

LLMs may require additional knowledge about processes and databases to implement PM tasks, for example, in anomaly detection and crafting accurate database queries. Some strategies are used to equip LLMs with this additional domain knowledge \cite{DBLP:journals/corr/abs-2309-00900}, including \emph{fine-tuning} and \emph{prompt engineering}.

\section{Evaluating LLMs in Process Mining}
\label{sec:discussion}

This section introduces criteria for selecting LLMs that are suitable for PM tasks. Moreover, we introduce criteria for evaluating their outputs. First, in \autoref{sec:llmsCapabilitiesForPm}, we discuss the fundamental capabilities needed for PM (long context window, acceptance of visual prompts, coding, factuality). Then, we introduce in \autoref{sec:relevantLlmsBenchmarks} general-purpose and process-mining-specific benchmarks to measure the different LLMs on process-mining-related tasks. To foster the development of process-mining-specific benchmarks and to be able to evaluate a given output, we propose in \autoref{sec:llmsOutputsEvaluation} different methods to evaluate the output of an LLM.

\vspace{-2mm}
\subsection{LLMs Capabilities Needed for Process Mining}
\label{sec:llmsCapabilitiesForPm}

\newlength{\taskcolwidth}
\setlength{\taskcolwidth}{3cm}
\newlength{\supersetcolwidth}
\setlength{\supersetcolwidth}{\dimexpr(\textwidth-\taskcolwidth)/9\relax}
\newlength{\supersetcolwidthtwo}
\setlength{\supersetcolwidthtwo}{\dimexpr(\textwidth-\taskcolwidth)/8\relax}

In this section, we discuss four important capabilities of LLMs for PM tasks:

\begin{itemize}
\item \noindent \emph{Long Context Window}:
Event logs in PM often include a vast amount of cases and events, challenging the \emph{context window} limit of LLMs, which restricts the token count in a prompt \cite{DBLP:journals/corr/abs-2401-01325}. Moreover, also the textual specification of process models requires a significant amount of information. The context window limit can be severe in many currently popular LLMs.\footnote{\tiny \url{https://community.openai.com/t/are-the-full-8k-gpt-4-tokens-available-on-chatgpt/237999}} Even simple abstractions like the ones introduced in  \cite{DBLP:journals/corr/abs-2307-12701} (directly-follows graph, list of process variants) may exceed this limitation. The context window, which is set during model training, must be large enough for the data size. Recent efforts aim to extend this limit, though quality may decline \cite{DBLP:journals/corr/abs-2401-01325,DBLP:journals/corr/abs-2309-00071}.
\item \noindent \emph{Accepting Visual Prompts}: Visualizations in PM, such as the dotted chart and the performance spectrum \cite{DBLP:conf/bpm/KlijnF19}, summarizing process behavior, empower analysts to spot interesting patterns not seen in tables. Interpreting visual prompts is key for semi-automated PM. Large Visual Models (LVMs) use architectures similar to language models trained on annotated image datasets \cite{DBLP:journals/corr/abs-2307-00855}. They perform tasks like object detection and image synthesis, recognizing patterns, textures, shapes, colors, and spatial relations.\footnote{GPT-4 and Google Bard/Gemini are popular models supporting both visual and textual prompts.}
\item \noindent \emph{Coding (Text-to-SQL) Capabilities}: With the context window limit preventing full event log inclusion in prompts, generating scripts and database queries is crucial for analyzing event data. As discussed in \autoref{sec:implStrategiesPMonLLMs}, text-to-SQL assists in filtering and analyzing event data. Key requirements for text-to-SQL in PM include understanding database schemas, performing complex joins, using database-specific operators (e.g., for calculating date differences), and translating PM concepts into queries.
Overall, modern LLMs offer excellent coding capabilities \cite{DBLP:journals/corr/abs-2307-12701}.
\item \noindent \emph{Factuality}: LLM hallucination involves generating incorrect or fabricated information \cite{DBLP:conf/emnlp/RawteCPSTCSD23}. Factuality measures an LLM's ability to cross-check its outputs against real facts or data, crucial for PM tasks like anomaly detection and root cause analysis. This may involve leveraging external databases \cite{DBLP:journals/corr/abs-2306-08302}, knowledge bases, or internet search \cite{DBLP:journals/corr/abs-2308-11432} for validation. For instance, verifying the sequence Cancel Order'' followed by Deliver Order'' against public data in anomaly detection. LLMs with web browsing can access up-to-date information, enhancing factuality.\footnote{\tiny \url{https://cointelegraph.com/news/chat-gpt-ai-openai-browse-internet-no-longer-limited-info-2021}}
\end{itemize}

\vspace{-2mm}
\subsection{Relevant LLMs Benchmarks}
\label{sec:relevantLlmsBenchmarks}

After identifying the required capabilities for LLMs in PM, benchmarking strategies are essential to measure the quality of
the textual outputs returned by the LLMs satisfying such capabilities.

Considering the wide array of available benchmarks for assessing LLMs behavior, we focus on identifying those most relevant to PM capabilities. In \cite{DBLP:journals/corr/abs-2307-03109}, a comprehensive collection of benchmarks is introduced. This section aims to select and utilize some of these benchmarks to evaluate various aspects of LLMs' performance in PM contexts.

\begin{itemize}
\item \noindent \emph{Traditional benchmarks}: Textual prompts are crucial for LLMs evaluation in PM. Benchmarks like AGIEval assess models via standardized exams  \cite{DBLP:journals/corr/abs-2304-06364}, and MT-Bench focuses on conversational and instructional capabilities \cite{DBLP:journals/corr/abs-2306-05685}. Another benchmark evaluates LLMs on prompts of long size \cite{DBLP:journals/corr/abs-2309-13345}.
\item \noindent \emph{Domain knowledge benchmarks}: Domain knowledge is essential for LLMs in PM to identify anomalies using metrics and context. Benchmarks like XIEZHI assess knowledge across different fields (economics, science, engineering) \cite{DBLP:journals/corr/abs-2306-05783}, while ARB evaluates expertise in areas like mathematics and natural sciences \cite{DBLP:journals/corr/abs-2307-13692}.
\item \noindent \emph{Visual benchmarks}: Understanding PM visualizations, such as dotted charts, is essential (c.f. \autoref{sec:llmsCapabilitiesForPm}). LLMs must accurately process queries on these visualizations. MMBench tests models on image tasks \cite{DBLP:journals/corr/abs-2307-06281}, and MM-Vet assesses recognition, OCR, among others \cite{DBLP:journals/corr/abs-2308-02490}. Yet, they may not fully meet PM visualization analysis needs, particularly in evaluating line orientations and point size/color.
\item \noindent \emph{Benchmarks for Text-to-SQL}: In PM, generating SQL from natural language is key for tasks like event log filtering. Benchmarks such as SPIDER and SPIDER-realistic test LLMs on text-to-SQL conversion \cite{DBLP:journals/corr/abs-2204-00498}. The APPS benchmark evaluates broader code generation abilities \cite{DBLP:conf/nips/HendrycksBKMAGB21}.
\item \noindent \emph{Fairness benchmarks}: they evaluate LLM fairness in PM by analyzing group treatment and bias detection. DecodingTrust measures LLM trustworthiness, covering toxicity, bias, robustness, privacy, ethics, and fairness \cite{DBLP:journals/corr/abs-2306-11698}.
\item \noindent \emph{Benchmarking the generation of hypotheses}: LLMs' ability to generate hypotheses from event data is vital to implement semi-autonomous PM agents. While specific benchmarks for hypothesis generation are lacking, related studies like \cite{tong2023automating} and \cite{DBLP:journals/corr/abs-2309-02726} evaluate LLMs using scientific papers.
\end{itemize}

In \autoref{tab:implBenchmLLMs}, we link process mining (PM) tasks to implementation paradigms and benchmarks. We discuss these tasks:

\begin{itemize}
\item \emph{Process description} requires understanding technical terms relevant to the domain, crucial for accurately describing processes.
\item \emph{Process modeling} involves generating models from text, using SQL for declarative and BPMN XML for procedural models. LLMs should offer various model hypotheses.
\item \emph{Anomaly detection} and \emph{root cause analysis} need domain knowledge to analyze process sequences or identify event attribute combinations causing issues.
\item \emph{Fairness} involves detecting biases by analyzing event attributes and values, necessitating hypothesis generation by LLMs.
\item \emph{Explaining and interpreting visualizations} requires extracting features from images and texts, offering contextual insights, like interpreting performance spectrum visualization \cite{DBLP:conf/bpm/KlijnF19}.
\item \emph{Process improvement} entails suggesting text proposals or new constraints to enhance current models, leveraging code generation capabilities and understanding process limitations.
\end{itemize}

\begin{table*}[!t]
\centering
\caption{Implementation paradigms and benchmarks for LLMs in the context of different PM tasks.}
\resizebox{0.75\textwidth}{!}{
\begin{tabular}{|l|*{3}{>{\centering\arraybackslash}p{\supersetcolwidth}|}*{6}{>{\centering\arraybackslash}p{\supersetcolwidth}|}}
\hline
\textbf{Task} & \multicolumn{3}{c|}{\textbf{Paradigms}} & \multicolumn{6}{c|}{\textbf{Benchmarks Classes}} \\
\cline{2-10}
 & \rotatebox{90}{Direct Provision} & \rotatebox{90}{Code Generation} & \rotatebox{90}{Hypotheses Generation} & \rotatebox{90}{Traditional} & \rotatebox{90}{Domain Knowledge} & \rotatebox{90}{Visual Prompts} & \rotatebox{90}{Text-to-SQL} & \rotatebox{90}{Fairness} & \rotatebox{90}{Hypotheses Generation} \\
\hline
Process Description & X &  &  & X & X &  &  &  &  \\
\hline
Process Modeling & X & X & X & X & X &  & X &  & X \\
\hline
Anomaly Detection & X &  & X & X & X &  & X & & X \\
\hline
Root Cause Analysis & X & & X & X & X &  & X &  & X \\
\hline
Ensuring Fairness & X &  & X & X & X &  & X & X & X \\
\hline
Expl. and Interpreting & X &  &  &  & X & X &  &  &  \\
Visualizations & ~ & ~ & ~ & ~ & ~ & ~ & ~ & ~ & ~ \\
\hline
Process Improvement & X & X & X & X & X &  & X & X & X \\
\hline
\end{tabular}
}
\label{tab:implBenchmLLMs}
\vspace{-3mm}
\end{table*}

While general-purpose benchmarks are already developed and are easily accessible, they are not entirely suited for the task of PM-on-LLMs. In particular, visual capabilities (explaining and interpreting PM visualizations) and autonomous hypotheses generation require more PM-specific benchmarks. However, little research exists on PM-specific benchmarks \cite{DBLP:journals/corr/abs-2307-12701,DBLP:journals/corr/abs-2307-02194}.

\vspace{-2mm}
\subsection{How to Evaluate LLMs Outputs}
\label{sec:llmsOutputsEvaluation}

This section outlines criteria for assessing the quality of outputs generated by LLMs in PM tasks, serving two primary objectives. The first objective is to assist users in identifying and addressing hallucinations and inaccuracies in LLMs' outputs.
The second aim is to establish criteria for developing an extensive benchmark specifically tailored to PM applications of LLMs. The strategies follow:

\begin{itemize}
\item \noindent \emph{Automatic evaluation} is particularly suited for text-to-SQL tasks. In this context, the formal accuracy and conciseness (indicated by the length of the produced query) of the SQL queries generated can be efficiently assessed. Additionally, the creation of declarative constraints, designed to enhance process execution, can also be evaluated in terms of their formal correctness.
\item \noindent \emph{Human evaluation} is essential for LLM tasks like direct querying and hypothesis generation. For direct querying tasks such as anomaly detection and root cause analysis, important criteria are \emph{recall} (the model's ability to identify expected insights) and \emph{precision} (the correctness of insights). These criteria also apply to hypothesis generation. Additionally, evaluating the feedback cycle's effectiveness in validating original hypotheses is crucial for these tasks.
\item \noindent \emph{Self-evaluation} in LLMs tackles hallucinations, as noted by \cite{DBLP:conf/emnlp/RawteCPSTCSD23}. Techniques include \emph{chain-of-thought}, where LLMs detail their reasoning, enhancing explanations \cite{DBLP:conf/nips/Wei0SBIXCLZ22}. \emph{Confidence scores} let LLMs assess their insights' reliability, discarding uncertain outputs for quality \cite{DBLP:journals/corr/abs-2309-16145}. \emph{Ensembling}, or using results from multiple LLM sessions, increases accuracy via majority voting or confidence checks \cite{DBLP:journals/corr/abs-2311-08692}. \emph{Self-reflection}, an LLM reviewing its or another's output, detects errors \cite{DBLP:journals/corr/abs-2312-09300}. In anomaly detection, using confidence scores to exclude doubtful anomalies and ensembling to confirm detections across sessions improves reliability.
\end{itemize}

\vspace{-2mm}
\section{Conclusion}
\label{sec:conclusion}

This paper examines LLM applications in PM, offering three main contributions: identification of necessary LLM capabilities for PM, review of benchmarks from literature, and strategies for evaluating LLM outputs in PM tasks. These strategies aim to build confidence in LLM use and establish benchmarks to assess LLM effectiveness across PM implementations.

Our discussion centers on current generative AI capabilities within PM, anticipating advancements like deriving event logs from videos. Despite future enhancements, the criteria discussed here should remain pertinent. Benchmarking for PM tasks on large language models (LLMs) will evolve, including both general and PM-specific benchmarks, yet the foundational aspects and methodologies are expected to stay consistent.

\bibliographystyle{splncs04}
\bibliography{references}

\end{document}